\documentclass[twocolumn,10pt,twoside]{IEEEtran}

\usepackage{cite}
\usepackage{url}
\usepackage{graphicx}
\usepackage{epstopdf}
\usepackage[cmex10]{amsmath}
\usepackage{amsfonts}
\usepackage{amssymb}
\usepackage{amsthm}
\usepackage{subfigure}
\usepackage{esint}
\usepackage{setspace}
\usepackage{algorithm}
\usepackage{color}

\begin{document}

\title{Delay-Optimal Biased User Association in Heterogeneous Networks}
\author{Fancheng~Kong, Xinghua~Sun,~\IEEEmembership{Member,~IEEE}, Victor~C.~M.~Leung,~\IEEEmembership{Fellow,~IEEE}, and Hongbo~Zhu
\thanks{F. Kong, X. Sun, and H. Zhu are with the Jiangsu Key Laboratory of Wireless Communications, College of Telecommunications and Information
Engineering, Nanjing University of Posts and Telecommunications, Nanjing 210003, China (e-mail: kfcshimaidi@163.com; xinghua.sun.cn@ieee.org; zhb@njupt.edu.cn).

V.~C.~M. Leung is with the Department of Electrical and Computer Engineering, University of British Columbia, Vancouver, BC V6T 1Z4, Canada (e-mail: vleung@ece.ubc.ca).}
}
\maketitle

\begin{abstract}
  In heterogeneous networks (HetNets), load balancing among different tiers can be effectively achieved by a biased user association scheme with which each user chooses to associate with one base station (BS) based on the biased received power. In contrast to previous studies where a BS always has packets to transmit, we assume in this paper that incoming packets intended for all the associated users form a queue in the BS. In order to find the delay limit of the network to support real-time service, we focus on the delay optimization problem by properly tuning the biasing factor of each tier. By adopting a thinned Poisson point process (PPP) model to characterize the locations of BSs in the busy state, an explicit expression of the average traffic intensity of each tier is obtained. On that basis, an optimization problem is formulated to minimize a lower bound of the network mean queuing delay. By showing that the optimization problem is convex, the optimal biasing factor of each tier can be obtained numerically.
  When the mean packet arrival rate of each user is small, a closed-form solution is derived.
  The simulation results demonstrate that the network queuing performance can be significantly improved by properly tuning the biasing factor.
  It is further shown that the network mean queuing delay might be improved at the cost of a deterioration of the network signal-to-interference ratio (SIR) coverage, which indicates a performance tradeoff between real-time and non-real-time traffic in HetNets.
\end{abstract}


\IEEEpeerreviewmaketitle
\section{Introduction}\label{I}
With widespread use of portable devices such as smart phones and tablets, cellular networks are facing an exponential growth of mobile data traffic \cite{Andrews}.
Meanwhile, real-time applications such as video chat and online gaming become more and more popular, which imposes stringent delay requirements on the network.
To deal with this ever-growing demand and high service requirement, micro base stations (BSs) such as pico and femto BSs are deployed to undertake the traffic pressure of macro BSs.
The network architecture is thus evolving to more dense and irregular heterogeneous networks (HetNets) \cite{Andrews2}.

Among all the techniques used in HetNets, load balancing plays a key role to determine the network performance. For example, if the traditional maximum downlink received power association scheme is adopted, users would tend to connect to macro BSs with a high transmission power. Macro BSs may thus easily become overloaded.
To purposely push users to micro BSs, a simple and efficient approach called the biased association scheme was proposed in \cite{Carlos}, with which each user assigns a biased value to the measured received power from BSs of each tier, and associates with the BS that has the largest mean biased received power.
Ye \emph{et al.} demonstrated in \cite{Ye} that the user's long-term rate can be greatly improved by carefully tuning the biasing factor.
With fixed locations of BSs and users, no tractable expression of a performance metric such as the signal-to-interference-plus-noise ratio (SINR) or rate coverage can be derived, and the optimal biasing factor could only be found empirically.
Stochastic geometry was then adopted in \cite{Dhillon,Singh,Singh2,Jo,Kong,Lin} to characterize the spatial distributions of BSs and users, and to quantify the average performance metric of the network.
For example, a Poisson point process (PPP) was adopted in \cite{Singh,Singh2} to represent the irregular deployment of the BSs. The optimal biasing factor for BSs of each tier was obtained therein to maximize the rate coverage. With a similar PPP model, the biasing factor was optimized in \cite{Lin,Kong} by maximizing the logarithm of the mean user rate.

\subsection{Related Works and Motivations}
Previous studies \cite{Ye,Dhillon,Singh,Singh2,Jo,Kong,Lin} assumed that the BSs always have packets to transmit, which presents a worst case for the SINR and rate coverage.
In practice, the BS load could vary significantly over a day.
In particular, the amount of user service requests can drop dramatically during non-peak traffic hours. The BSs are thus more likely to be idle during such periods, but still consume energy \cite{Wu}.
On the other hand, due to a high deployment density, one micro BS would cover a small area.
The void cell problem \cite{Liu,Peng} then emerges, where some micro BSs don't have any associated users. Such BS thus solely acts as an interfering source.
To improve energy efficiency and reduce the interference, the techniques to selectively switch off a fraction of BSs according to the traffic load have attracted extensive attention \cite{Jung,Oh,Guo,Dhillon2,Cao}. For example, the authors in \cite{Oh} proposed a distributed on/off switching based algorithm in cellular networks to decide the minimum set of active BSs.
By arguing that a cellular BS could operate in normal mode, sleep mode, or expansion mode, Guo \emph{et al.} \cite{Guo} proposed a scheme that determines which mode the BS should choose based on the load condition, such that the energy consumption is minimized.
Dhillon \emph{et al.} \cite{Dhillon2} adopted a thinned-PPP model by assigning an active probability to BSs of each tier and thus characterized the network signal-to-interference ratio (SIR) coverage. With the same model, Cao \emph{et al.} \cite{Cao} derived the optimal BS density of each tier to minimize the network energy consumption under a certain rate constraint.

The aforementioned studies \cite{Jung,Oh,Guo,Dhillon2,Cao} focused on the traffic load variance over a large time scale, i.e., peak and non-peak hours, and did not consider queuing in each BS. Most of them aimed to improve the network energy efficiency.
One BS, nevertheless, can vary between busy and idle states over a small time scale due to the dynamic packet arrivals of its associated users, in which case the packet delay could be taken into account.
In practice, with the proliferation of real-time multimedia applications, the packet delay is becoming an important quality-of-service (QoS) metric.
For example, an end-to-end latency over 250 ms for real-time multimedia is generally considered to be unacceptable \cite{Tu}.
In the literature, there has been limited studies on the queuing analysis.
The queuing performance of a single cell was evaluated in \cite{Bonald} and \cite{Borst} in CDMA systems for the first time, based on which the packet blocking probability and the packet queuing delay were characterized. These studies focused on only one independent queue by assuming constant interference over the entire cell.
In HetNets, nevertheless, BSs of various types are deployed with higher densities, and the queuing behavior of one BS is closely related with others, leading to the coupled-queue problem \cite{Dhillon2}.
Specifically, whether a BS transmits or not will affect the interference level experienced by other BSs that occupy the same spectrum resources, and will consequently determine the service rates of these co-channel BSs.
The service rate in turn affects the chance that one BS has packets to transmit or not.

The analysis of coupled queues is a long-standing open problem, and even solving a special case of two interacting queues is challenging \cite{Adan}.
Zhuang \emph{et al.} \cite{Zhuang} modeled multiple interacting queues as a continuous-time Markov chain (CTMC) with fixed BS locations in the network.
By minimizing the average packet queuing delay, the optimal spectrum allocation pattern was obtained for each BS.
Similarly, Cheng \emph{et al.} \cite{Cheng} optimized the average queuing delay of network subjected to the BS power constraints.
They formulated a Markov decision process (MDP) problem based on the instantaneous channel state information and queue state information, and proposed an adaptive user scheduling and power control policy. However, the state space of the Markov process may become huge as the network scales up, and the analysis would become intractable. Hence, this motivates us to deal with the coupled queue problem with the tool of stochastic geometry to account for the random BS deployment, and derive the mean performance metrics analytically such that some insights can be gained for system design.

\subsection{Our Approaches and Contributions}
In this paper, we consider a $K$-tier HetNet where users and BSs of all tiers are randomly distributed, i.e., follow a PPP distribution.
Similar to previous studies \cite{Dhillon,Singh,Singh2,Jo,Kong,Lin}, it is assumed that each user adopts a biased association scheme to choose one BS with the maximum biased received power.
In contrast to previous studies \cite{Ye,Dhillon,Singh,Singh2,Jo,Kong,Lin,Jung,Oh,Guo,Dhillon2,Cao}, we consider that the packet requests from the users form a queue in their associated BSs.
The traffic intensity of one BS thus varies with the aggregate packet requests of all its associated users.
To simplify the analysis, frequency partitioning across tiers is assumed in this paper.
Although the queues of BSs from different tiers are not correlated, the queuing performance of one BS would affect the experienced interference of other co-channel BSs in the same tier.
To decouple the queuing behavior of BSs in the same tier, we resort to the approximation of replacing each BS's individual traffic intensity with the average traffic intensity of its tier.
The spatial distribution of BSs in the busy state can thus be approximately characterized by a thinned-PPP model \cite{Dhillon2}.
The SIR coverage of each tier is then obtained, based on which an explicit expression of the average traffic intensity of each tier is further derived, and is shown to be an increasing function of the biasing factor of each tier.

In order to find the delay limit that the network can achieve, an optimization problem is formulated to minimize a lower bound of the network mean queuing delay by optimizing over the biasing factor of each tier based on the derived average traffic intensity.
It is shown that the optimization problem is convex, and the optimal biasing factor can be numerically obtained.
When the mean packet arrival rate of each user is small, an explicit expression of the optimal biasing factor of each tier is obtained. With equal bandwidth allocation across tiers, it is further shown that each user should associate with its nearest BS.
A case study of a 2-tier HetNet shows that the optimal biasing factor is sensitive to the bandwidth allocation of each tier.
To characterize the network capacity to support non-real-time services, the network SIR coverage is further derived.
The contributions of this paper are summarized as follows.
\begin{itemize}
  \item By assuming queuing in each BS, an explicit expression of the average traffic intensity of each tier is derived, which is shown to be an increasing function of the biasing factor of each tier.
  \item An optimization problem of a lower bound of the network mean queuing delay is formulated, and is shown to be convex with respect to the biasing factor of each tier. When the mean packet arrival rate of each user is small, an explicit solution is obtained.
  \item Simulation results of a 2-tier case demonstrate that the network mean queuing delay can be significantly reduced by properly tuning the biasing factor of each tier. In the meanwhile, a tradeoff is revealed between the network mean queuing delay and the network SIR coverage, which indicates that the service provider should strike a balance between the performance of real-time and non-real-time services.
\end{itemize}

The rest of this paper is organized as follows. The system model is presented in Section \ref{II}. An optimization problem to minimize a lower bound of the network mean queuing delay is formulated and studied in Section \ref{III}. A case study of a 2-tier HetNet is presented in Section \ref{IV}. Conclusions and future works are given in Section \ref{V}.
\section{System Model}\label{II}
Consider a $K$-tier heterogeneous network where BSs in the $k$th tier form an independent PPP $\Phi_k$ with an intensity of $\lambda_k$, $k\in\{1,...,K\}$.
Users, on the other hand, form another independent homogeneous PPP $\Phi_\textrm{u}$ with an intensity of $\lambda_\textrm{u}$ over the whole network.
Frequency partitioning across tiers is assumed in this paper. In particular, BSs of the same tier share the spectrum with a bandwidth of $W_k$, $k\in\{1,...,K\}$, and BSs of different tiers occupy non-overlapping frequency bands.
Therefore, for each user in the downlink, the associated BS acts as a desired signal transmitter, and other BSs of the same tier are interfering sources.
Consider a typical user located at the origin.
Denote the distance between this typical user and a Tier-$k$ BS as $x_k$, and the transmission power of a Tier-$k$ BS as $P_k$. The received power $P_\textrm{R}$ for a typical user from this BS can then be written as
\begin{equation} \label{RecPower}
{P_\textrm{R}} = {P_k}{g_k}{x^{-\alpha}_k},
\end{equation}
where $g_k$ is a small-scale fading coefficient, which is assumed to follow an i.i.d exponential distribution of unit mean, i.e., ${g_k}{\sim} \exp\{1\}$, and $\alpha$ is the path-loss coefficient, which is assumed to be the same for all BSs in the network. Note that shadowing, i.e., log-normal fading, can be modeled by the randomness of the BSs' and users' locations \cite{Blaszczyszyn}.

In this paper, we consider a biased association scheme where users associate with one BS according to the maximum mean biased received power \cite{Ye,Dhillon,Singh,Singh2,Jo,Kong,Lin}. In particular, for a typical user located at the origin, it measures the mean received power from each tier's BSs, and chooses a Tier-$k$ BS if
\begin{equation}\label{BiasedScheme}
{P_k}{B_k}x_{k,\min }^{ - \alpha }\ge{P_j}{B_j}x_{j,\min }^{ - \alpha }\;\;\;\forall j\in\{1,...,K\},
\end{equation}
where ${B_j}$ denotes the biasing factor of Tier $j$ and $x_{j,\min }$ is the distance between the user and the nearest Tier-$j$ BS.

For each user in the network, assume that its packet requests follow an independent Poisson process with a mean arrival rate $\gamma$, and the packet length is exponentially distributed with mean $L$. The incoming packets for all users form a queue in the associated BS, and the BS will transmit these packets in a first-in-first-serve (FIFS) fashion.
To avoid users in poor channel conditions occupying the BS, we consider a fixed rate modulation and coding format.
In particular, a BS will serve a user only when its instantaneous SIR exceeds a threshold $\tau$, and will drop its packet request otherwise.
Note that due to a high BS deployment intensity, the background noise is dominated by the interference, and is therefore ignored in this paper.
According to Shannon's formula, the service rate for each user that is associated to a Tier-$k$ BS can be obtained as
\begin{equation}\label{ServiceRate}
{\mu}_k  = \frac{{{W_k}}}{L}{\log _2}\left( {1 + \tau } \right).
\end{equation}
For a randomly selected Tier-$k$ BS, ${\textrm{BS}}_{k,i}$, its traffic intensity, ${\rho}_{k,i}$, can be obtained as
\begin{equation}\label{TrafficIntensity}
{\rho}_{k,i}=\frac{{\gamma}_{k,i}}{{\mu}_{k}},
\end{equation}
where ${\gamma}_{k,i}$ is the mean aggregate packet arrival rate of all its associated users.
Note that ${\rho}_{k,i}$ can also be interpreted as the busy probability or the utilization of ${\textrm{BS}}_{k,i}$ when ${\rho}_{k,i}\leq1$.
Due to a varied association region, each BS has a different mean aggregate packet arrival rate and thus has a different traffic intensity.
\section{Queuing Delay}\label{III}
In this section, we will formulate an optimization problem of the network mean queuing delay, which is an important performance metric of QoS.
As the mean queuing delay of a BS increases with a higher busy probability, we will first characterize the average traffic intensity of each tier's BSs, $\rho_k$.
\subsection{Average Traffic Intensity of Tier-$k$ BSs, $\rho_k$}\label{IIIA}
For a randomly selected Tier-$k$ ${\textrm{BS}}_{k,i}$, since it delivers a packet only when the SIR exceeds a certain threshold $\tau$, its mean aggregate packet arrival rate can be obtained as
\begin{align}\label{RealArrivalRate}
{\gamma _{k,i}} = \gamma {N_{k,i}}{\rm{Pr}\left[\rm{SIR}_{k,i}>\tau\right]}
\end{align}
where $N_{k,i}$ is the number of users that are associated to ${\textrm{BS}}_{k,i}$ and ${\rm{Pr}\left[\rm{SIR}_{k,i}>\tau\right]}$ denotes the SIR coverage of ${\textrm{BS}}_{k,i}$, i.e., the probability that the SIR of a random user associated to ${\textrm{BS}}_{k,i}$ is larger than the threshold $\tau$.
By combining \eqref{ServiceRate}-\eqref{RealArrivalRate}, the average traffic intensity of Tier-$k$ BSs can be obtained as
\begin{align}
{\rho _k} = E\left[ {{\rho _{k,i}}} \right] = E\left[ {\frac{{\gamma {N_{k,i}}{\rm{Pr}\left[\rm{SIR}_{k,i}>\tau\right]}}}{{{\mu _k}}}} \right] \nonumber
\end{align}
\begin{align}\label{BusyProb}
&= \frac{\gamma }{{{\mu _k}}}  E\left[ {{N_{k,i}}} \right]  E\left[ {\rm{Pr}\left[\rm{SIR}_{k,i}>\tau\right]} \right] \nonumber\\
&=\frac{{\gamma L}}{{{W_k}{{\log }_2}\left( {1 + \tau } \right)}} E\left[ {{N_{k}}} \right]  {\rm{P}}\left[ {{\rm{SIR}_k} > \tau } \right],
\end{align}
where $E\left[ {{N_{k}}} \right]$ denotes the average number of users associated with a Tier-$k$ BS and ${\rm{P}}\left[ {{\rm{SIR}_k} > \tau } \right]$ denotes the SIR coverage of all Tier-$k$ BSs, i.e., the probability that the SIR of a typical user associated with a Tier-$k$ BS exceeds the threshold $\tau$.
As the average traffic intensity, $\rho_k$, is determined by the average number of associated users, $E\left[ {{N_{k}}} \right]$, and the SIR coverage, ${\rm P}\left[ {{\rm SIR}_k > {\tau} } \right]$, we will characterize these two components in the following.

According to \cite{Jo}, the average number of users associated with a Tier-$k$ BS, $E\left[ {{N_{k}}} \right]$, has been obtained as
\begin{align}\label{AvNum}
E\left[{N_{k}}\right]=\frac{{{\lambda _\textrm{u}}{A_k}}}{{{\lambda _k}}},
\end{align}
where $A_k$ denotes the probability for a typical user to be associated with a Tier-$k$ BS. Note that the association probability $A_k$ has been derived in \cite{Jo} as
\begin{equation}\label{AssoProb}
{A_k} = \frac{{{\lambda _k}{{\left( {{P_k}{B_k}} \right)}^{2/\alpha }}}}{{\sum\limits_{j = 1}^K {{\lambda _j}{{\left( {{P_j}{B_j}} \right)}^{2/\alpha }}} }} =\frac{1}{\sum\limits_{j = 1}^K {\widetilde{\lambda}}_j \left({\widetilde{B}}_j {\widetilde{P}}_j\right)^{2/\alpha}},
\end{equation}
where ${\widetilde{\lambda}}_j={\lambda _j}/{\lambda _k}$, ${\widetilde{P}}_j={P_j}/{P _k}$, and ${\widetilde{B}}_j={B_j}/{B _k}$ denote the normalized intensity, the normalized transmission power, and the normalized biasing factor of Tier $j$, respectively, conditioned on Tier $k$ being a serving tier.

Recall that BSs of Tier $k$ form a PPP $\Phi_k$ with an intensity of $\lambda_k$. Moreover, for a randomly selected ${\textrm{BS}}_{k,i}$ where $i\in\Phi_k$, the traffic intensity ${\rho}_{k,i}$ can be interpreted as its busy probability when ${\rho}_{k,i}\leq 1$.
The set of Tier-$k$ BSs being busy, therefore, forms a thinned point process $\Phi'_k\subseteq\Phi_k$ by including ${\textrm{BS}}_{k,i}\in\Phi_k$ with the probability ${\rho}_{k,i}$ \cite{Stoyan}.
Since the traffic intensity of one BS is different from each other, the thinned point process $\Phi'_k$ is non-homogeneous.
To simplify the analysis, it can be approximately viewed as a homogeneous PPP with intensity
\begin{equation}\label{ThinnedIntensity}
\lambda'_k={\rho_{k}} {\lambda_k}.
\end{equation}
It will be demonstrated in Section \ref{IV} that the average traffic intensity achieves a good approximation.
For a typical user that is associated with a Tier-$k$ BS, the interference all comes from busy BSs of the same tier. According to \eqref{RecPower}, the SIR of this typical user can then be written as
\begin{equation}\label{SIR}
{\rm SIR}_k = \frac{{{P_k}{g_{{x_{k,0}}}}x^{ - \alpha }_{k,0}}}{{\sum\limits_{j \in {{\Phi'_k}\backslash {\textrm{BS}}_{k,0}}} {{P_k}{g_{k,j}}{x^{ - \alpha }_{k,j}}} }},
\end{equation}
where $x_{k,0}$ and ${x_{k,j}}$ denote the distance from the typical user to the associated BS ${\textrm{BS}}_{k,0}$ and the $j$th interfering Tier-$k$ BS, respectively; $g_{k,0}$ and ${g_{k,j}}$ denote the small-scale fading coefficient of ${\textrm{BS}}_{k,0}$ and the $j$th interfering Tier-$k$ BS, respectively.
In \eqref{SIR}, ${\textrm{BS}}_{k,0}$ and ${{\Phi'_k}\backslash {\textrm{BS}}_{k,0}}$ denote the associated Tier-$k$ BS of this typical user and the set of interfering Tier-$k$ BSs, respectively. Note that as frequency partitioning is assumed across tiers, there is no inter-tier interference, and the interfering sources consist of all the busy Tier-$k$ BSs besides the associated ${\textrm{BS}}_{k,0}$.
Following a similar approach in \cite{Singh}, we have the following lemma that presents the SIR coverage of a Tier-$k$ BS.
\vspace{6pt}

\noindent \textbf{Lemma 1.} \emph{The SIR coverage of a Tier-$k$ BS can be written as}
\begin{equation}\label{SIRCoverage}
{\rm P}\left[ {{\rm SIR}_k > \tau } \right] = \frac{1}{{{A_k}{\rho _k}Z\left( {\tau ,\alpha ,1} \right) + 1}},
\end{equation}
\emph{where $Z\left( {\tau ,\alpha ,1} \right) = {\tau ^{\frac{2}{\alpha }}}\int_{{{\left( {1/\tau } \right)}^{\frac{2}{\alpha} }}}^\infty  {\frac{{du}}{{1 + {u^{\frac{\alpha}{2} }}}}}$, and the probability $A_k$ for a typical user to be associated with a Tier-$k$ BS is given in \eqref{AssoProb}.}
\vspace{6pt}

\noindent According to Lemma 1, the outage probability of Tier $k$, ${\rm P}\left[ {{\rm SIR}_k \leq \tau } \right]$, can be written as ${\rm P}\left[ {{\rm SIR}_k \leq \tau } \right]=\frac{{A_k}{\rho _k}Z\left( {\tau ,\alpha ,1} \right)}{{{A_k}{\rho _k}Z\left( {\tau ,\alpha ,1} \right) + 1}}$.
If Tier-$k$ BSs are always busy, i.e., $\rho_k =1$, the outage probability ${\rm P}\left[ {{\rm SIR}_k \leq \tau } \right]$ reduces to [6, Eq. (17)].

By combining \eqref{BusyProb}, \eqref{AvNum}, and \eqref{SIRCoverage}, the average traffic intensity $\rho_k$ of Tier-$k$ BSs can be derived as
\begin{align} \label{rho_k}
{\rho _k} = {\frac{{ {-} {\lambda _k}{R_k} + {{\left[ {{{\left( {{\lambda _k}{R_k}} \right)}^2} + 4\gamma {\lambda _\textrm{u}}{\lambda _k}A_k^2{R_k}LZ} \right]}^{\frac{1}{2}}}}}{{2{A_k}{\lambda _k}{R_k}Z}}}
\end{align}
where
\begin{align} \label{Rate}
{R_k} = {W_k}{\log _2}\left( {1 + \tau } \right),
\end{align}
and $Z$ denotes $Z\left( {\tau ,\alpha ,1} \right)$ for simplicity.
It is indicated in \eqref{rho_k} that $\rho_k$ is critically determined by the mean packet arrival rate of each user $\gamma$ and the association probability $A_k$. It is clear that $\rho_k$ increases as $\gamma$ increases. On the other hand, the following lemma presents the monotonicity of the average traffic intensity $\rho_k$ of Tier-k BSs with respect to the association probability $A_k$.

\vspace{6pt}
\noindent \textbf{Lemma 2.} \emph{The average traffic intensity $\rho_k$ of Tier-$k$ BSs is an increasing function of its association probability, $A_k$.}
\vspace{6pt}
\begin{proof}
See Appendix B.
\end{proof}
\vspace{6pt}

\noindent Intuitively, as the probability of a user being associated with a Tier-$k$ BS increases, more users from other tiers will be offloaded to BSs of Tier $k$, i.e., $E\left[ {{N_{k}}} \right]$ becomes larger, which leads to an increment of the traffic intensity.

To this end, an explicit expression of $\rho_k$ has been derived in \eqref{rho_k}. When the mean packet arrival rate of each user $\gamma$ is small, the average traffic intensity $\rho_k$ of Tier-k BSs can be approximately written as
\begin{align}
&{\rho _k} = \frac{{ {-} 1 + \left[{1 + 4\gamma {\lambda _\textrm{u}}A_k^2{{\left( {{\lambda _k}{R_k}} \right)}^{ - 1}}LZ}\right]^{\frac{1}{2}}}}{{2{A_k}Z}}\nonumber
\end{align}
\begin{align}\label{rho_app}
&\mathop \approx \limits^{(a)} \frac{{ - 1 + {{ {1 + 2\gamma {\lambda _\textrm{u}}A_k^2{{\left( {{\lambda _k}{R_k}} \right)}^{ - 1}}LZ} }}}}{{2{A_k}Z}}\;\;\;\;\; \nonumber \\
& =\frac{{\gamma {\lambda _\textrm{u}}L{A_k}}}{{{\lambda _k}{R_k}}},\;\;\;\;\;
\end{align}
where (a) follows from the fact that
\begin{align}\label{Taylor2}
\left[1 + \frac{4\gamma {\lambda _\textrm{u}}A_k^2LZ}{{\lambda _k}{R_k}}\right]^{\frac{1}{2}}\approx { {1 + \frac{{2\gamma {\lambda _\textrm{u}}A_k^2}LZ}{{{ {{\lambda _k}{R_k}} }}} }}.
\end{align}
\subsection{Queuing Delay Optimization}\label{IIIB}
\newcounter{mytempeqncnt2}
\begin{figure*}[!t]
\normalsize
\setcounter{mytempeqncnt2}{\value{equation}}
\setcounter{equation}{17}
\begin{align} \label{DelayTotal}
\bar D = \sum\limits_{k = 1}^K \frac{{{\lambda _k}}}{{{\sum\limits_{j = 1}^K \lambda _j}}}\cdot \bar D_k=\sum\limits_{k = 1}^K {\frac{{2{A_k}\lambda _k^2LZ}}{{\sum\limits_{j = 1}^K {{\lambda _j}} \left( {2{A_k}{\lambda _k}Z{R_k} + {\lambda _k}{R_k} - {{\left[ {{{\left( {{\lambda _k}{R_k}} \right)}^2} + 4\gamma {\lambda _u}{\lambda _k}A_k^2{R_k}LZ} \right]}^{\frac{1}{2}}}} \right)}}}
\end{align}
\setcounter{equation}{\value{mytempeqncnt2}}
\hrulefill
\vspace*{4pt}
\end{figure*}

\newcounter{mytempeqncnt1}
\begin{figure*}[!t]
\normalsize
\setcounter{mytempeqncnt1}{\value{equation}}
\setcounter{equation}{24}
\begin{align} \label{Patial}
\frac{{\partial \bar D}}{{\partial {A_k}}} &= 2{\lambda _K}Z  \frac{{{R_K}{{\left( {1 - \sum\limits_{j = 1}^{K - 1} {{A_j}} } \right)}^{ - 2}} - R_K^2{{\left( {1 - \sum\limits_{j = 1}^{K - 1} {{A_j}} } \right)}^{ - 3}}{{\left[ {R_K^2{{\left( {1 - \sum\limits_{j = 1}^{K - 1} {{A_j}} } \right)}^{ - 2}} + 4\gamma {\lambda _\textrm{u}}\lambda _K^{ - 1}{R_K}LZ} \right]}^{ - \frac{1}{2}}}}}{{{{\left\{ {2Z{R_K} + {R_K}{{\left( {1 - \sum\limits_{j = 1}^{K - 1} {{A_j}} } \right)}^{ - 1}} - {{\left[ {R_K^2{{\left( {1 - \sum\limits_{j = 1}^{K - 1} {{A_j}} } \right)}^{ - 2}} + 4\gamma {\lambda _\textrm{u}}\lambda _K^{ - 1}{R_K}LZ} \right]}^{\frac{1}{2}}}} \right\}}^2}}}\nonumber\\&- 2{\lambda _k}Z  \frac{{{R_k}A_k^{ - 2} - R_k^2A_k^{ - 3}{{\left( {R_k^2A_k^{ - 2} + 4\gamma {\lambda _\textrm{u}}\lambda _k^{ - 1}{R_k}LZ} \right)}^{ - \frac{1}{2}}}}}{{{{\left[ {2Z{R_k} + {R_k}A_k^{ - 1} - {{\left( {R_k^2A_k^{ - 2} + 4\gamma {\lambda _\textrm{u}}\lambda _k^{ - 1}{R_k}LZ} \right)}^{\frac{1}{2}}}} \right]}^2}}}=0,\;\;\;\; k\in\{1,\ldots,K{-}1\}.
\end{align}
\setcounter{equation}{\value{mytempeqncnt1}}
\hrulefill
\vspace*{4pt}
\end{figure*}

In this section, we will further minimize a lower bound of the network mean queuing delay based on the average traffic intensity by optimally tuning the biasing factors of all tiers.
As each BS can be modeled as a M/D/1 queuing system, the mean queuing delay $D_k$ of Tier $k$ BSs can be obtained as
\begin{equation}\label{DelayTierk}
{D_k} = E\left[\frac{L}{{R_k\left( {1 - {\rho _{k,i}}} \right)}}\right].
\end{equation}
Since \eqref{DelayTierk} is difficult to characterize, we resort to its lower bound using the convexity of $1/(1-\rho_{k,i})$, i.e., we have
\begin{equation}\label{DelayTierkLowerBound}
D_k\ge  {\bar D_{k}}= \frac{L}{{R_k\left( {1 - E\left[{\rho _{k,i}}\right]} \right)}}=\frac{L}{{R_k\left( {1 - \rho_k} \right)}}.
\end{equation}
By combining \eqref{ServiceRate}, \eqref{rho_k} and \eqref{DelayTierkLowerBound}, the lower bound of the mean queuing delay of the whole network $\bar D$ can then be written as \eqref{DelayTotal}, which is shown on the top of next page.

It can be observed from \eqref{DelayTotal} that the lower bound of the mean queuing delay $\bar D$ is critically determined by the association probability $A_k$. To minimize $\bar D$, we have the following optimization problem
\setcounter{equation}{18}
\begin{subequations}\label{DelayOptimizationp}
\begin{equation}\label{Problemp}
\bar D^* = \mathop {{\rm{minimize}}}\limits_{{\{ {A_k}\} _{\forall k{\in}\{1,\ldots,K\}}}} {\rm{ }}\bar D,
\end{equation}
\begin{equation}\label{Constraints1p}
\rm{s.t.}\sum\limits_{k = 1}^K {{A_k}}  = 1,
\end{equation}
\begin{align}\label{Constraints2p}
\rho_k<1,\;k\in\{1,\ldots,K\}.
\end{align}
\end{subequations}
\noindent Note that instead of directly optimizing over the biasing factor of each tier, we optimize over the association probabilities $\{A_k\}_{\forall k}$ in \eqref{DelayOptimizationp} to obtain the optimal solution $\{A^*_k\}_{\forall k}$.
The optimal normalized biasing factor of Tier $k$ conditioned on Tier $i$, $\{{\widetilde{B}}^*_k\}_{\forall k}$, can then be readily obtained as
\begin{align}\label{AssoProbtoBF}
{\widetilde{B}}^*_k=\frac{P_i {\left(\lambda_i A^*_k\right)}^{\frac{\alpha}{2}}}{P_k {\left(\lambda_k A^*_i\right)}^{\frac{\alpha}{2}}},\;k\in\{1,...,K\},
\end{align}
according to \eqref{AssoProb}.
On the other hand, the constraint \eqref{Constraints1p} comes from the fact that each user should associate with one BS, and the constraint \eqref{Constraints2p} ensures that the lower bound of the network's mean queuing delay is bounded, which leads to the following lemma.

\vspace{6pt}
\noindent \textbf{Lemma 3.}
\emph{For the lower bound $\bar D_{k}$, when the mean packet arrival rate $\gamma  > \frac{{\left( {Z {+} 1} \right){\lambda _k}{R_k}}}{{{\lambda _\textrm{u}}L}}$, it is bounded if the association probability}
\begin{equation}\label{UnsaturatedCondition}
0<{A_k} < \frac{{{\lambda _k}{R_k}}}{{\gamma {\lambda _\textrm{u}}L - {\lambda _k}{R_k}Z}};
\end{equation}
\emph{otherwise, it will become unbounded.}
\emph{When $\gamma  < \frac{{\left( {Z {+} 1} \right){\lambda _k}{R_k}}}{{{\lambda _\textrm{u}}L}}$, it is always bounded.}
\vspace{6pt}
\begin{proof}
See Appendix C.
\end{proof}
\vspace{6pt}

According to Lemma 3, constraint \eqref{Constraints2p} can be further written as
\begin{align}\label{Constraints2pr}
\Bigg\{ {\begin{array}{*{20}{c}}
{0 < {A_k} < \frac{{{\lambda _k}{R_k}}}{{\gamma {\lambda _\textrm{u}}L {-} {\lambda _k}{R_k}Z}}},&{\gamma  > \frac{{\left( {Z + 1} \right){\lambda _k}{R_k}}}{{{\lambda _u}L}}}\\
{0 < {A_k} < 1},&{\gamma  < \frac{{\left( {Z + 1} \right){\lambda _k}{R_k}}}{{{\lambda _\textrm{u}}L}}}
\end{array}},
\end{align}
where $k\in\{1,\ldots,K\}$.
First note that \eqref{Constraints2pr} does not have a feasible region if and only if
\begin{subequations}\label{judge}
\begin{align}\label{judge1}
\gamma >\mathop {\max }\limits_{\forall k} \left\{ {\frac{{\left( {Z {+} 1} \right){\lambda _k}{R_k}}}{{{\lambda _\textrm{u}}L}}} \right\}
\end{align}
and
\begin{align}\label{judge2}
\sum\limits_{k = 1}^K {\frac{{{\lambda _k}{R_k}}}{{\gamma {\lambda _\textrm{u}}L - {\lambda _k}{R_k}Z}}}  < 1,
\end{align}
\end{subequations}
according to \eqref{Constraints2pr}.
Intuitively, when the mean packet arrival rate of each user $\gamma$ is too large, \eqref{Constraints2pr} can be written as $0 < {A_k} < \frac{{{\lambda _k}{R_k}}}{{\gamma {\lambda _\textrm{u}}L {-} {\lambda _k}{R_k}Z}}$ for each Tier $k$, $k\in\{1,\ldots,K\}$, which leads to $\sum\limits_{k = 1}^K {{A_k}}  < 1$ according to \eqref{judge2}. In this case, the lower bound of the network mean queuing delay will always be unbounded.
If \eqref{judge} does not hold, the feasible region of the optimization problem \eqref{DelayOptimizationp} can be further written as
\begin{align}
&\textbf{A}{=}\Bigg\{\left( {{A_1},...,{A_{K{-}1}}} \right),\left| 0<{A_j} <\min\Big\{1, \frac{{{\lambda _j}{R_j}}}{{\gamma {\lambda _\textrm{u}}L {-} {\lambda _j}{R_j}Z}}\Big\}, \right.  \nonumber
\end{align}
\begin{align}\label{region}
&j \in {\{1...,K {-} 1} \};\; \max\Big\{0,1 {-} \frac{{{\lambda _K}{R_K}}}{{\gamma {\lambda _\textrm{u}}L {-} {\lambda _K}{R_K}Z}}\Big\}<\sum\limits_{j = 1}^{K {-} 1} {{A_j}}\nonumber \\ &  <1 \Bigg\},
\end{align}
where $A_K$ is eliminated according to the constraint \eqref{Constraints1p} without loss of generality.
It is shown in Appendix D that the objective function \eqref{Problemp} is convex within the feasible region $\textbf{A}$.
Nevertheless, there may not exist a solution in $\textbf{A}$ through solving \eqref{Patial}, which is shown at the top of this page, by setting the partial derivative of $\bar D$ with respect to the association probability $A_k$ to zero.
The following lemma rules out this possibility and guarantees that the optimal association probabilities $\{A^*_k\}_{\forall k}$ can always be obtained by finding the solution of \eqref{Patial} within $\textbf{A}$.

\vspace{6pt}
\noindent \textbf{Lemma 4.}
\emph{\eqref{Patial} has a unique solution within the feasible region $\textbf{A}$, which is the optimal association probabilities $\{A^*_k\}_{\forall k}$.}
\vspace{6pt}
\begin{proof}
See Appendix E.
\end{proof}
\vspace{6pt}

So far we have demonstrated how to find the optimal association probability of each tier $A^*_k$ by solving \eqref{Patial} numerically. Recall that it is indicated in Lemma 3 that when the mean packet arrival rate of each user $\gamma <\mathop {\min }\limits_{\forall k} \left\{ {\frac{{\left( {Z {+} 1} \right){\lambda _k}{R_k}}}{{{\lambda _\textrm{u}}L}}} \right\}$, we have the average traffic intensity $\rho_k<1$ for all tiers, and the lower bound of the mean queuing delay $\bar D_k$ is always bounded for each tier. In this case, the average traffic intensity $\rho_k$ is simply written as \eqref{rho_app}, and an explicit optimal association probability $A^*_k$ for each tier can be obtained, which is shown in the following lemma.


\vspace{6pt}
\noindent \textbf{Lemma 5.}
\emph{When the mean packet arrival rate of each user $\gamma <\mathop {\min }\limits_{\forall k} \left\{ {\frac{{\left( {Z {+} 1} \right){\lambda _k}{R_k}}}{{{\lambda _\textrm{u}}L}}} \right\}$, the optimal association probability of Tier $k$ $A^*_k$ to minimize the lower bound of the network mean queuing delay $\bar D$ can be written as}
\setcounter{equation}{25}
\begin{equation}\label{Optimal}
{A^*_k} = \frac{{{\lambda _k}}}{{\sum\limits_{j = 1}^K {{\lambda _j}} }} + \frac{{{\lambda _k}{\log _2}\left( {1 + \tau } \right)\sum\limits_{j = 1}^K {{\lambda _j}(W_k-W_j)} }}{{\gamma {\lambda _\textrm{u}}L\sum\limits_{j = 1}^K {{\lambda _j}} }}.
\end{equation}
\vspace{6pt}

\noindent The detailed derivation can be found in Appendix F.

Intuitively, if the bandwidth of Tier $k$ is larger than that of Tier $j$, i.e., $W_k>W_j$, the service rate of Tier $k$ will be larger, indicating a better queuing performance. Therefore, the Tier-$k$ BSs will undertake more traffic from other tiers by having a larger association probability.
With equal bandwidth allocation among all tiers, i.e., $W_i=W_j$, $i,k\in\{1,\ldots,K\}$, the optimal association probability of a Tier-$k$ BS can be further written as
\begin{equation}
A^*_k=\frac{{{\lambda _k}}}{{\sum\limits_{j = 1}^K {{\lambda _j}} }}
\end{equation}
according to \eqref{Optimal}. The corresponding optimal normalized biasing factor ${\widetilde{B}}^*_k$ of Tier $k$, conditioned on Tier $i$, is thus given by
\begin{equation}\label{OptimalEQBW}
{\widetilde{B}}^*_k {=} \frac{1}{\widetilde{P}_k},
\end{equation}
where $\widetilde{P}_k$ is the normalized transmission power of Tier $k$ conditioned on Tier $i$.
It is indicated in \eqref{OptimalEQBW} that in this case, each user chooses the nearest BS.
The traffic load is thus evenly distributed among all BSs, which leads to similar queuing performance with the same service rate of each tier's BSs.

\begin{algorithm}[t]
\caption{Procedure to optimize the association probability when the mean packet arrival rate of each user $\gamma <\mathop {\min }\limits_{\forall k} \left\{ {\frac{{\left( {Z {+} 1} \right){\lambda _k}{R_k}}}{{{\lambda _\textrm{u}}L}}} \right\}$}
  1: \textbf{Input:} $\lambda_k$, $W_k$ for each tier, and other system parameters\\ $\lambda_\textrm{u}$, $L$, $\gamma$, $\tau$.\\
  2: \textbf{Initialize:} a set of index $C=\{1,\ldots,K\}$ where optimal\\ association probability of Tier $k$ is not determined.\\
  3: Calculate the solution set $\{A^*_k\} _{\forall k{\in}C}$ by \eqref{Optimal}.\\
  4: \textbf{for} $\forall k\in C$, construct a set $S=\left\{ {m\left| {A^*_m < 0,\forall m \in C} \right.} \right\}$.\\
  5: \textbf{if} $S=\emptyset$, \textbf{return} $\{A^*_k\} _{\forall k{\in}C}$.\\
  6: \textbf{else}, \textbf{for} $\forall m\in S$, let $\lambda_m = 0$ and $A^*_m = 0$, delete $m$ from $C$.\\
  7: \textbf{end if}\\
  8: \textbf{go to} Step 3.\\
\end{algorithm}

Note from \eqref{Optimal} that if there exists one tier, say Tier $m$, such that $\log_2\left(1{+}\tau\right)\sum\limits_{j = 1}^K {{\lambda _j}( {{ R_m} {-} {R_j}} )}<{-}\gamma \lambda_u L$, and then we have $A^*_m<0$.
To minimize the lower bound of the network mean queuing delay, the association probability of Tier $m$ should be close to zero.
Intuitively, if the bandwidth of Tier $m$ is much smaller than that of other tiers, then few users should associate with Tier-$m$ BSs due to the low service rate.
In this case, the association probability $A_m$ could then be simply set as $A_m =0$, i.e., Tier-$m$ BSs are turned off.
The procedure to obtain the optimal association probability when $\gamma <\mathop {\min }\limits_{\forall k} \left\{ {\frac{{\left( {Z {+} 1} \right){\lambda _k}{R_k}}}{{{\lambda _\textrm{u}}L}}} \right\}$ is summarized in Algorithm 1.

\section{Case Study}\label{IV}
In this section we will demonstrate the analytical results in Section \ref{III} by simulations of a 2-tier HetNet. The base stations and the users are drawn from PPPs with high intensities, and the background noise is ignored in the simulations. This setting, for example, can correspond to a dense heterogeneous network that consists of macro cellular BSs and micro Wi-Fi access points, each of which uses a non-overlapping frequency band. Each point of the simulation results is obtained by averaging all the BSs on a time scale of $10^5\;\rm{s}$.
The system parameters used in simulations are summarized in Table I.
\begin{table}
\label{Table}\normalsize
\
\caption{Simulation Parameters}
\centering
\begin{tabular}{lll}
  \hline
  Parameter & Value \\
  \noalign{\global\arrayrulewidth1pt}\hline\noalign{\global\arrayrulewidth0.4pt}
  User Density ${\lambda _\textrm{u}}$ & $10^{-2}\;\rm{m}^{{-}2}$ \\
  Tier-1 BS Density $\lambda_1$ & $10^{-4}\;\rm{m}^{{-}2}$ \\
  Tier-2 BS Density $\lambda_2$ & $5{*}10^{-4}\;\rm{m}^{{-}2}$ \\
  Tier-1 BS Transmission Power $P_1$ & 46\;dBm\\
  Tier-2 BS Transmission Power $P_2$ & 35\;dBm\\
  Path Loss Coefficient $\alpha$ & 4\\
  Mean Packet Length $L$ & 0.1\;Mb\\
  \hline
\end{tabular}
\end{table}
\begin{figure}[tbp]
\centering
\includegraphics[width=85mm,height=75mm]{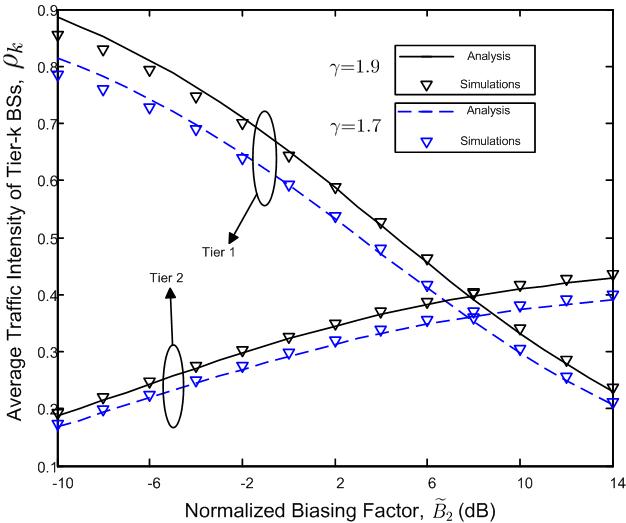}
\caption{Average traffic intensity of each tier $\rho_k$ versus the normalized biasing factor ${\widetilde{B}}_2$ with various values of the mean packet arrival rate of each user $\gamma$. $W_1=10\rm{MHz}$, $W_2=6\rm{MHz}$, and $\tau=1$.}
\label{Prob}
\end{figure}

Fig. \ref{Prob} illustrates how the average traffic intensity of each tier, i.e., $\rho_1$ and $\rho_2$, varies with the normalized biasing factor ${\widetilde{B}}_2$ with various values of the mean packet arrival rate of each user $\gamma$.
It can be observed from Fig. \ref{Prob} that the average traffic intensity of Tier 1, $\rho_1$, decreases as the normalized biasing factor ${\widetilde{B}}_2$ increases, while that of Tier 2, $\rho_2$, increases.
Intuitively, since the association probability of a Tier-2 BS, $A_2$, increases as the normalized biasing factor ${\widetilde{B}}_2$ increases according to \eqref{AssoProb}, more users that associate with Tier-1 BSs would be offloaded to Tier-2 BSs, which leads to an increment of $\rho_2$ according to Lemma 2.
Moreover, due to a larger deployment intensity of the Tier-2 BSs, the users that originally associate with only one Tier-1 BS can be offloaded to several neighboring Tier-2 BSs.
Hence, the decline rate of $\rho_1$ is larger than the increasing rate of $\rho_2$.
It can be clearly seen from Fig. \ref{Prob} that the simulation results match with the analysis well with a wide range of the normalized biasing factor, indicating that replacing each BS's traffic intensity by the average traffic intensity in \eqref{ThinnedIntensity} achieves a good approximation.
\begin{figure}[tbp]
\centering
\includegraphics[width=85mm,height=75mm]{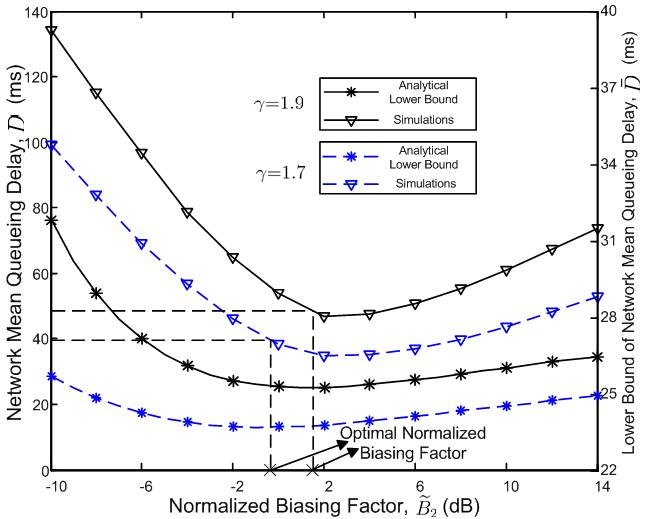}
\caption{Network mean queuing delay $D$ and its lower bound $\bar D$ versus the normalized biasing factor ${\widetilde{B}}_2$ with various values of the mean packet arrival rate of each user $\gamma$. $W_1=10\rm{MHz}$, $W_2=6\rm{MHz}$, and $\tau=1$.}
\label{Delay}
\end{figure}

Fig. \ref{Delay} further demonstrates how the network mean queuing delay $D$, as well as its lower bound $\bar D$, vary with the normalized biasing factor ${\widetilde{B}}_2$.
For the sake of comparison, the y-axis on the left hand side of Fig. \ref{Delay} denotes the network mean queuing delay $D$ while on the right hand side it denotes the lower bound $\bar D$.
To obtain the network mean queuing delay in simulations, BSs that have an unbounded queuing delay are not taken account of.
It can be observed from Fig. \ref{Delay} that the trend of the network mean queuing delay $D$ resembles that of its lower bound $\bar D$. Both $D$ and $\bar D$ are very sensitive to the normalized biasing factor ${\widetilde{B}}_2$.
If ${\widetilde{B}}_2$ is not carefully tuned, the delay performance could be greatly degraded.
For example, when $\gamma=1.9$, the network mean queuing delay $D$ is as high as 135 ms with the normalized biasing factor ${\widetilde{B}}_2=-10\;\rm{dB}$, which is not acceptable to many delay-sensitive applications.
Moreover, due to a similar trend between the network mean queuing delay $D$ and its lower bound $\bar D$, the optimal normalized biasing factor of $\bar D$ is close to that of $D$.
Therefore, by properly tuning the normalized biasing factor ${\widetilde{B}}_2$ according to \eqref{AssoProbtoBF} and \eqref{Patial}, the mean queuing delay performance can be improved significantly.
With the mean packet arrival rate of each user $\gamma=1.9$, for instance, the optimal normalized biasing factor is obtained as $\widetilde{B_2^*}=1.7\;\rm{dB}$, and the corresponding network mean queuing delay $D$ can be reduced to be 48 ms.

\begin{figure}[tbp]
\centering
\includegraphics[width=85mm,height=80mm]{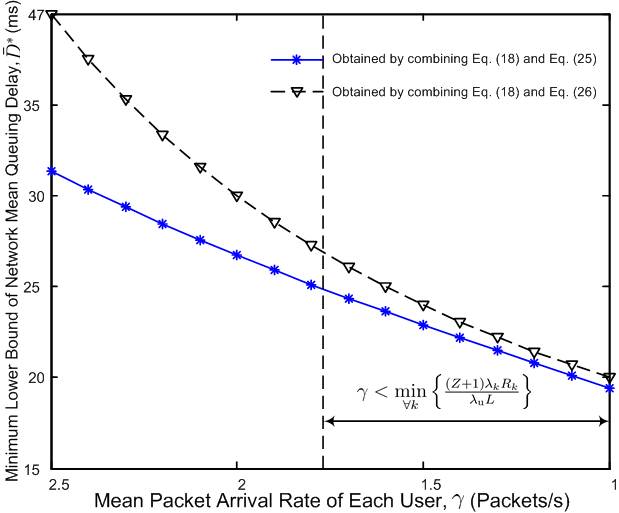}
\caption{Minimum lower bound of the network mean queuing delay $\bar D^*$ versus the mean packet arrival rate of each user $\gamma$. $W_1=10\rm{MHz}$, $W_2=6\rm{MHz}$, and $\tau=1$.}
\label{Fig3}
\end{figure}

\begin{figure}[tbp]
\centering
\subfigure[]{
\includegraphics[width=85mm,height=75mm]{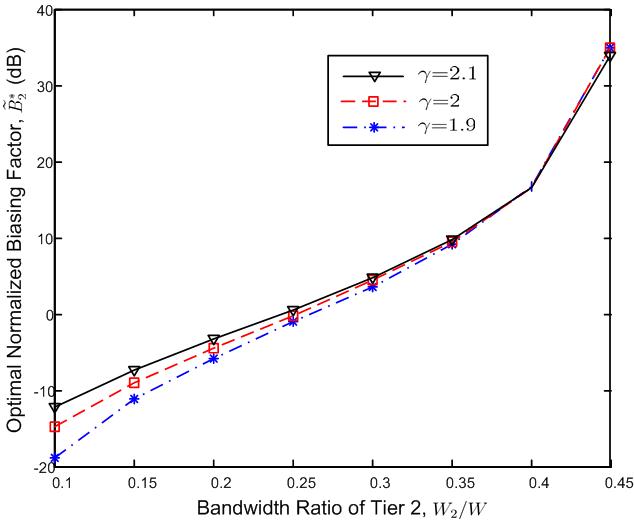}
\label{OptimalBF}}
\subfigure[]{
\includegraphics[width=85mm,height=77mm]{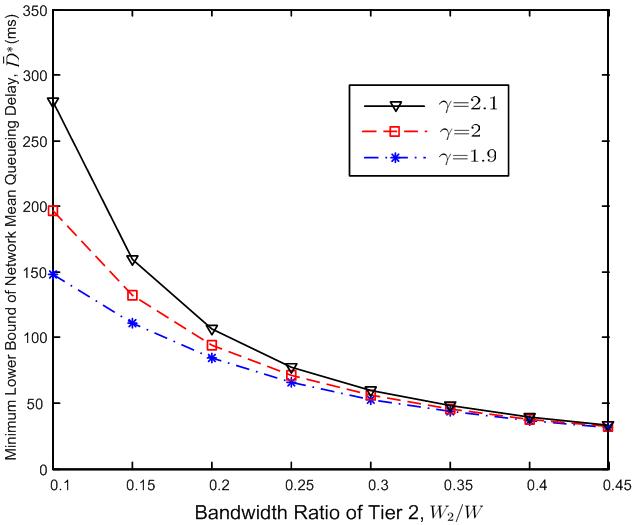}
\label{OptimalDelay}}
\caption{Optimal normalized biasing factor $\widetilde{B_2^*}$ and the minimum lower bound of the network mean queuing delay $\bar D^*$ versus the bandwidth ratio of Tier 2 $W_2/W$ with various values of the mean packet arrival rate of each user $\gamma$. $W{=}12\rm{MHz}$ and $\tau=1$. (a) Optimal normalized biasing factor $\widetilde{B_2^*}$. (b) Minimum lower bound of the network mean queuing delay $\bar D^*$.}
\label{OptimalRes}
\end{figure}

Recall that it is indicated in Lemma 5 that when the mean packet arrival rate satisfies $\gamma <\mathop {\min }\limits_{\forall k} \left\{ {\frac{{\left( {Z {+} 1} \right){\lambda _k}{R_k}}}{{{\lambda _\textrm{u}}L}}} \right\}$, the minimum lower bound of the network mean queuing delay $\bar D^*$ can be obtained by combining \eqref{DelayTotal} and \eqref{Optimal}.
Fig. \ref{Fig3} further compares the minimum lower bound of the network mean queuing delay $\bar D^*$ obtained by combining \eqref{DelayTotal} and \eqref{Patial} with that by combining \eqref{DelayTotal} and \eqref{Optimal}, respectively.
It can be observed from Fig. \ref{Fig3} that the gap between the two curves diminishes as the mean packet arrival rate of each user $\gamma$ decreases.
When $\gamma <\mathop {\min }\limits_{\forall k} \left\{ {\frac{{\left( {Z {+} 1} \right){\lambda _k}{R_k}}}{{{\lambda _\textrm{u}}L}}} \right\}$, the minimum lower bound of the network mean queuing delay $\bar D^*$ obtained by combining \eqref{DelayTotal} and \eqref{Optimal} is quite close to that obtained by combining \eqref{DelayTotal} and \eqref{Patial}.
When the mean packet arrival rate of each user $\gamma$ is large, i.e., $\gamma \geq \mathop {\min }\limits_{\forall k} \left\{ {\frac{{\left( {Z {+} 1} \right){\lambda _k}{R_k}}}{{{\lambda _\textrm{u}}L}}} \right\}$, there is a large gap between the curves in Fig. \ref{Fig3}. Therefore, the optimal association probabilities $\{A^*_k\}_{\forall k}$ should be instead obtained by numerically solving \eqref{Patial}. As Lemma 4 guarantees, \eqref{Patial} has a unique solution within the feasible region \textbf{A}, which is the optimal association probability $\{A^*_k\}_{\forall k}$.

\begin{figure}[tbp]
\centering
\subfigure[]{
\includegraphics[width=85mm,height=67mm]{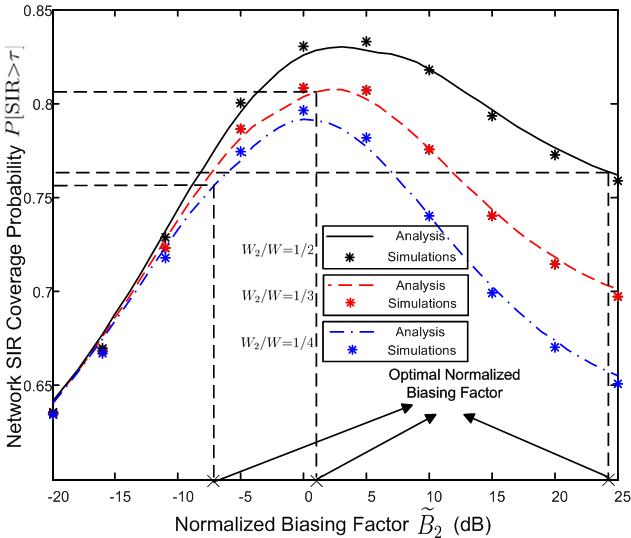}
\label{CoverageProb1}}
\subfigure[]{
\includegraphics[width=85mm,height=67mm]{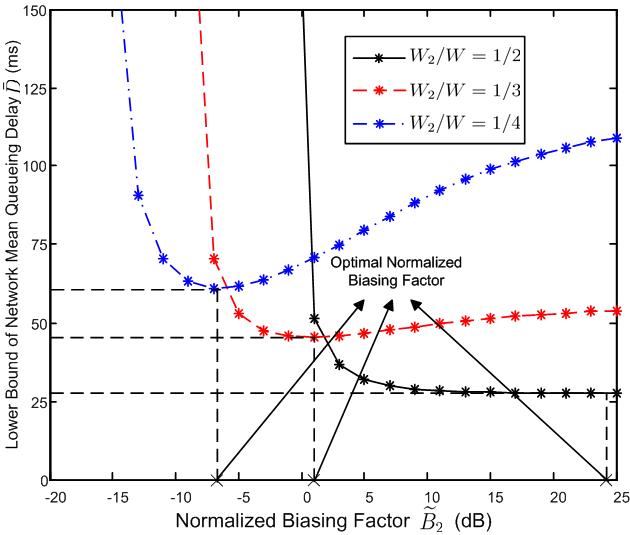}
\label{CoverageProb2}}
\subfigure[]{
\includegraphics[width=85mm,height=67mm]{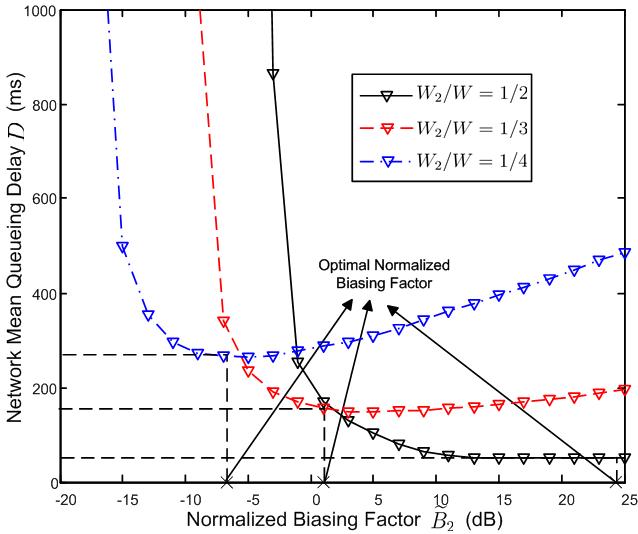}
\label{CoverageProb3}}
\caption{The network SIR coverage and the network mean queuing delay performance with various bandwidth ratios of Tier 2 $W_2/W$. $\gamma=1.8$, $W=12\rm{MHz}$, and $\tau=1$. (a) Network SIR coverage ${\rm P}\left[ {{\rm SIR} > \tau } \right]$. (b) Network mean queuing delay $D$. (c) Lower bound of the network mean queuing delay $\bar D$.}
\label{CoverageProb}
\end{figure}

Fig. \ref{OptimalRes} further illustrates how the optimal normalized biasing factor, $\widetilde{B_2^*}$, and the corresponding minimum lower bound of the network mean queuing delay, $\bar D^*$, vary with the bandwidth ratio of Tier 2, $W_2/W$, with various values of the mean packet arrival rate of each user $\gamma$.
Note that the total bandwidth $W=W_1+W_2$ is fixed here.
It can be observed from Fig. \ref{OptimalBF} that for a given $\gamma$, the optimal normalized biasing factor $\widetilde{B_2^*}$ increases as $W_2/W$ increases. Intuitively, as the bandwidth of Tier 2, $W_2$, increases, Tier-2 BSs can provide a higher service rate to the associated users. The optimal $\widetilde{B}^*_2$ should thus become larger so as to encourage more users to be associated with Tier-2 BSs.
Moreover, it can be observed from Fig. \ref{OptimalBF} that as $W_2/W$ increases, the optimal normalized biasing factor $\widetilde{B}^*_2$ becomes insensitive to the mean packet arrival rate of each user $\gamma$.
The minimum lower bound of the network mean queuing delay $\bar D^*$, on the other hand, decreases as $W_2/W$ increases, as Fig. \ref{OptimalDelay} demonstrates.


While minimizing the network mean queuing delay is desirable for real-time traffic, the SIR coverage is an important performance metric to support non-real-time traffic for service providers.
According to \eqref{SIRCoverage}, the network SIR coverage ${\rm P}\left[ {{\rm SIR} > \tau } \right]$ can be written as
\begin{align}\label{TotalSIRCoverage}
&{\rm P}\left[ {{\rm SIR} > \tau } \right] = \sum\limits_{k = 1}^K A_k \cdot {\rm P}\left[ {{\rm SIR}_k > \tau } \right] \nonumber\\
&\;\;\;\;\;\;= \sum\limits_{k = 1}^K \frac{A_k}{{{A_k}{\rho _k}Z\left( {\tau ,\alpha ,1} \right) + 1}}.
\end{align}
Fig. \ref{CoverageProb1} demonstrates how the network SIR coverage ${\rm P}\left[ {{\rm SIR} > \tau } \right]$ varies with the normalized biasing factor ${\widetilde{B}}_2$ with various values of the bandwidth ratio $W_2/W$.
It can be observed from Fig. \ref{CoverageProb1} that there exists an optimal normalized biasing factor with which the network SIR coverage is maximized.
Intuitively, when ${\widetilde{B}}_2$ is too large, a large fraction of users that are originally associated with Tier-1 BSs are offloaded to Tier-2 BSs.
As these users are close to the interfering Tier-1 BSs and have long distances to their associated Tier-2 BSs, they have very poor channel conditions, which leads to a low SIR coverage of the network.
Similarly, when ${\widetilde{B}}_2$ is too small, the network SIR coverage also deteriorates.
In addition, it can be seen from Fig. \ref{CoverageProb1} that the optimal normalized biasing to maximize ${\rm P}\left[ {{\rm SIR} > \tau } \right]$ is insensitive to the bandwidth allocation. In the meanwhile, the optimal normalized biasing factor $\widetilde{B_2^*}$ to minimize $\bar D$ increases as $W_2/W$ increases, as illustrated in Fig. \ref{CoverageProb3} indicating a tradeoff between the network mean queuing delay and the network SIR coverage. For example, if $W_2/W=1/2$, the optimal normalized biasing factor is obtained as $\widetilde{B_2^*}=24\rm{dB}$, with which the network SIR coverage greatly deteriorates.
In this case, the service providers should properly tune the biasing factor in HetNets such that a desired point on the tradeoff curve can be achieved to balance the performances of real-time traffic and non-real-time traffic.

\begin{figure}[tbp]
\centering
\subfigure[]{
\includegraphics[width=85mm,height=68mm]{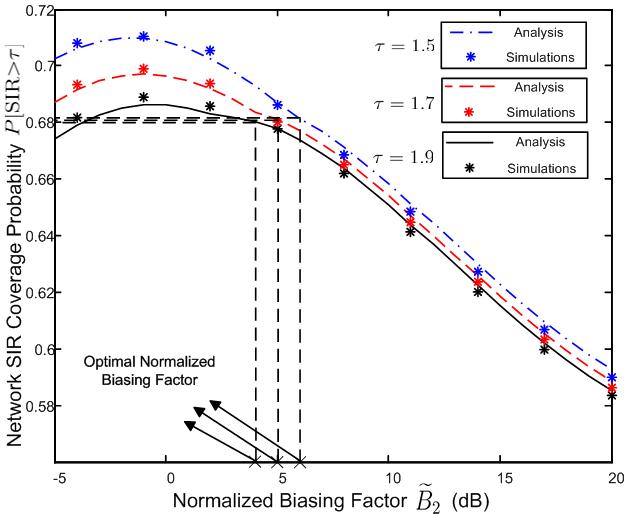}
\label{Fig6a}}
\subfigure[]{
\includegraphics[width=85mm,height=68mm]{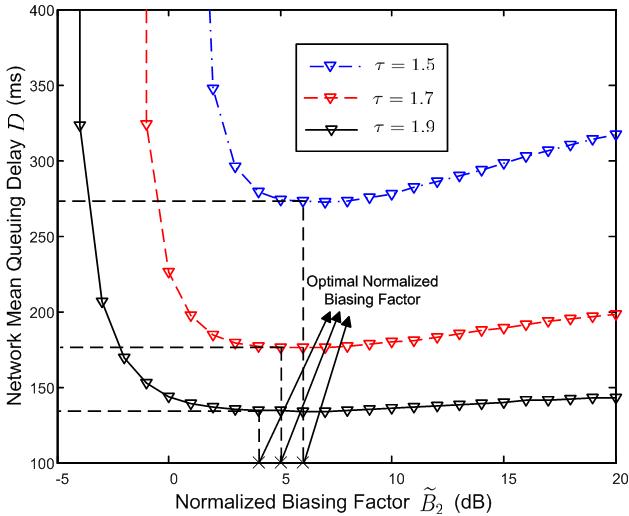}
\label{Fig6b}}
\subfigure[]{
\includegraphics[width=85mm,height=68mm]{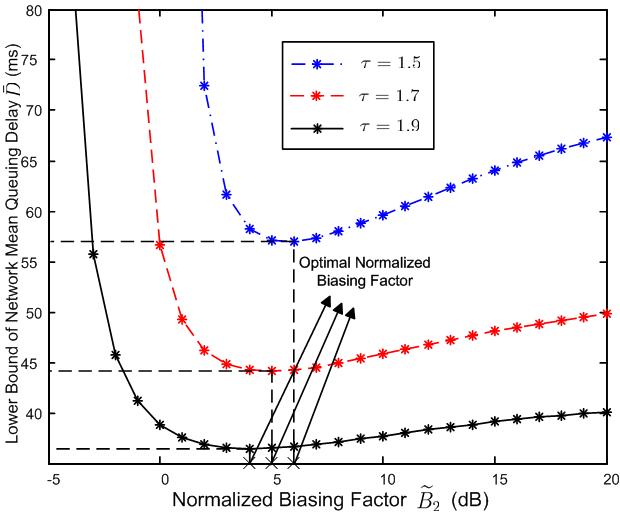}
\label{Fig6c}}
\caption{The network SIR coverage and the network queuing delay performance with various values of the SIR threshold $\tau$. $W_1=8\rm{MHz}$, $W_2=4\rm{MHz}$, and $\gamma=3.8$, (a) Network SIR coverage ${\rm P}\left[ {{\rm SIR} > \tau } \right]$. (b) Network mean queuing delay $D$. (c) Lower bound of the network mean queuing delay $\bar D$.}
\label{Fig6}
\end{figure}
As the SIR threshold $\tau$ critically determines the network mean queuing delay and the network SIR coverage, Fig. \ref{Fig6} further demonstrates the impact of the SIR threshold $\tau$ on these two performance metrics.
It can be observed from Fig. \ref{Fig6} that for a given normalized biasing factor ${\widetilde{B}}_2$, the network SIR coverage ${\rm P}\left[ {{\rm SIR} > \tau } \right]$ decreases as the SIR threshold $\tau$ increases. In the meanwhile, both the network mean queuing delay $D$ and its lower bound $\bar D$ decrease as $\tau$ increases. Intuitively, with a higher SIR threshold $\tau$, the mean aggregate packet arrival rate of each BS becomes lower while the service rate becomes higher, leading to a better queuing performance.
In addition, it is illustrated in Fig. \ref{Fig6a} that the optimal normalized biasing factor to maximize the network SIR coverage ${\rm P}\left[ {{\rm SIR} > \tau } \right]$ is insensitive to the SIR threshold $\tau$, while the optimal normalized biasing factor $\widetilde{B_2^*}$ to minimize $\bar D$ increases as $\tau$ decreases, as Fig. \ref{Fig6c} demonstrates. Intuitively, although the service rates of both macro and micro BSs become lower with a smaller $\tau$, macro BSs are more likely to become overloaded as their deployment density is much lower than that of micro BSs. The optimal normalized biasing factor $\widetilde{B_2^*}$ should thus become larger to undertake the load pressure from macro BSs.
By comparing Fig. \ref{Fig6a} with Fig. \ref{Fig6b} and Fig. \ref{Fig6c}, it can be found that with a smaller SIR threshold $\tau$, the deterioration of the network mean queuing delay $D$ becomes much more severe if the normalized biasing factor is optimally tuned to maximize the network SIR coverage ${\rm P}\left[ {{\rm SIR} > \tau } \right]$, indicating a more significant tradeoff between the network SIR coverage and the network mean queuing delay.

\section{Conclusion and Future Work}\label{V}
In this paper we have studied how to optimally tune the biasing factor of each tier in HetNets in order to minimize a lower bound of the network mean queuing delay.
It is shown that the network queuing performance can be significantly improved when the biasing factor of each tier is optimally tuned.
The characterization of the optimal biasing factor provides guidance for real-time service provisioning in HetNets.
The case study of a 2-tier HetNet further illustrates that the network mean queuing delay and the network SIR coverage might not be optimized simultaneously by tuning the biasing factor, indicating a performance tradeoff between real-time and non-real-time services.

It is worth mentioning that it is assumed in this paper that one BS will serve a user with a constant rate if its SIR exceeds a threshold.
In practice, nevertheless, the service rate could depend on the channel conditions. In this case, as the biasing factor of one tier decreases, the mean service rate of this tier increases as the users located at the edge of the cells are offloaded. The queuing performance of this tier can thus be improved due to a lower mean aggregate packet arrival rate and a higher mean service rate. Therefore, there would exist an optimal biasing factor for each tier such that the traffic load is balanced across tiers and the network mean queuing delay is minimized.
On the other hand, if the biasing factor of one tier is too large, the SIR coverage of this tier degrades, which would drag down the network SIR coverage.
Therefore, the network mean queuing delay may be optimized at the cost of the network SIR coverage.
The tradeoff between the network SIR coverage and the network mean queuing delay in this case is an interesting issue that needs further study.

In addition, it is assumed that orthogonal spectrum resources are allocated to different tiers. In practice, nevertheless, universal frequency reuse may be adopted so that all the other BSs may act as interfering sources for one BS.
Therefore, the average traffic intensities of different tiers would be correlated.
The characterization of the queuing performance under such circumstances deserves much attention in future study.
\begin{appendices}

\section{Proof of Lemma 2}
\begin{proof}
According to \eqref{rho_k}, the first-order derivative of the average traffic intensity $\rho_k$ with respect to $A_k$ is given by
\begin{align}\label{AppendixB1}
\frac{{d{\rho _k}}}{{d{A_k}}} = \frac{{4\gamma L{\lambda _\textrm{u}}\lambda _k^2R_k^2A_k^2{Z^2}{\Delta ^{ {-} \frac{1}{2}}} {-} {\lambda _k}{R_k}Z\left( { {-} {\lambda _k}{R_k} + {\Delta ^{\frac{1}{2}}}} \right)}}{{2{{\left( {{A_k}{\lambda _k}{R_k}Z} \right)}^2}}},
\end{align}
where $\Delta  = \lambda _k^2R_k^2 + 4\gamma {\lambda _\textrm{u}}{\lambda _k}{R_k} A_k^2 LZ$.
The numerator on the right hand side of \eqref{AppendixB1} can be further written as
\begin{align}
4\gamma L{\lambda _\textrm{u}}\lambda _k^2R_k^2A_k^2{Z^2}{\Delta ^{ {-} \frac{1}{2}}} - {\lambda _k}{R_k}Z\left( { {-} {\lambda _k}{R_k} + {\Delta ^{\frac{1}{2}}}} \right) \nonumber
\end{align}
\begin{align}\label{AppendixB2}
= \frac{{\lambda _k^2R_k^2Z\left[ {{{\left( {\lambda _k^2R_k^2 + 4\gamma {\lambda _\textrm{u}}{\lambda _k}{R_k}A_k^2 LZ} \right)}^{\frac{1}{2}}} - {\lambda _k}{R_k}} \right]}}{{{\Delta ^{\frac{1}{2}}}}} >0.
\end{align}
By combining \eqref{AppendixB1} and \eqref{AppendixB2}, we have $\frac{{d{\rho _k}}}{{d{A_k}}}>0$, which indicates that $\rho_k$ monotonically increases as $A_k$ increases.
\end{proof}

\section{Proof of Lemma 3}
\begin{proof}
It has been shown in Lemma 2 that the average traffic intensity $\rho_k$ monotonically increases as the association probability $A_k$ increases. With $A_k<1$, we then have
\begin{align}
\rho_k={\frac{{ {-} {\lambda _k}{R_k} + {{\left[ {{{\left( {{\lambda _k}{R_k}} \right)}^2} + 4\gamma {\lambda _\textrm{u}}{\lambda _k}{R_k}{A_k^2}LZ} \right]}^{\frac{1}{2}}}}}{{2{A_k}{\lambda _k}{R_k}Z}}}\nonumber
\end{align}
\begin{align}\label{AppendixC1}
<{\frac{{ {-} {\lambda _k}{R_k} + {{\left[ {{{\left( {{\lambda _k}{R_k}} \right)}^2} + 4\gamma {\lambda _\textrm{u}}{\lambda _k}{R_k}LZ} \right]}^{\frac{1}{2}}}}}{2{{\lambda _k}{R_k}Z}}}.
\end{align}
In the following, we divide the discussion into two parts:

$\left. 1 \right)$ If ${\frac{{ {-} {\lambda _k}{R_k} + {{\left[ {{{\left( {{\lambda _k}{R_k}} \right)}^2} + 4\gamma {\lambda _\textrm{u}}{\lambda _k}{R_k}LZ} \right]}^{\frac{1}{2}}}}}{2{{\lambda _k}{R_k}Z}}}<1$, i.e., $\gamma  < \frac{{\left( {Z {+} 1} \right){\lambda _k}{R_k}}}{{{\lambda _\textrm{u}}L}}$, we have
\begin{align}\label{AppendixC2}
\rho_k<1
\end{align}
according to \eqref{AppendixC1}. In this case, $\bar D_{k}$ will always be bounded if $\gamma  < \frac{{\left( {Z {+} 1} \right){\lambda _k}{R_k}}}{{{\lambda _\textrm{u}}L}}$.

$\left. 2 \right)$ If $\gamma  > \frac{{\left( {Z {+} 1} \right){\lambda _k}{R_k}}}{{{\lambda _u}L}},$ $\bar D_{k}$ will be bounded if and only if
\begin{equation}\label{AppendixC4}
{\frac{{ {-} {\lambda _k}{R_k} + {{\left[ {{{\left( {{\lambda _k}{R_k}} \right)}^2} + 4\gamma {\lambda _\textrm{u}}{\lambda _k}A_k^2{R_k}LZ} \right]}^{\frac{1}{2}}}}}{{2{A_k}{\lambda _k}{R_k}Z}}} < 1.
\end{equation}
Accordingly, we have
\begin{equation}\label{AppendixC5}
{{A_k} < \frac{{{\lambda _k}{R_k}}}{{\gamma {\lambda _\textrm{u}}L - {\lambda _k}{R_k}Z}}}.
\end{equation}
\end{proof}
\section{Proof of Convexity of \eqref{DelayOptimizationp}}
\begin{proof}
According to \eqref{DelayTierkLowerBound}, the second-order derivative of $\bar D_k$ with respect to $A_k$ can be written as
\begin{equation}\label{convex1}
\frac{{{d^2}{{\bar D}_k}}}{{d{A_k}^2}} = \frac{2L}{{{{R_k\left( {1 - {\rho _k}} \right)}^3}}} \cdot {\left( {\frac{{d{\rho _k}}}{{d{A_k}}}} \right)^2} + \frac{L}{{{{R_k\left( {1 - {\rho _k}} \right)}^2}}} \cdot \frac{{{d^2}{\rho _k}}}{{d{A_k}^2}}.
\end{equation}
Substituting \eqref{AppendixB1} into \eqref{convex1} yields
\begin{align}
&\frac{{{d^2}{{\bar D}_k}}}{{d{A_k}^2}} > \frac{L}{{{{R_k\left( {1 - {\rho _k}} \right)}^2}}} \cdot \left[ {2{{\left( {\frac{{d{\rho _k}}}{{d{A_k}}}} \right)}^2} + \frac{{{d^2}{\rho _k}}}{{d{A_k}^2}}} \right] \nonumber \\
& = \frac{L}{{{{R_k\left( {1 - {\rho _k}} \right)}^2}A_k^4{Z^2}\Delta }} \cdot \Bigg(4\gamma {\lambda _u}L\lambda _k^2R_k^2{Z^2}A_k^3 + 2\Delta \nonumber
\end{align}
\begin{align}\label{convex2}
&  \;\;\;\;\;+ 2{\lambda _k}{R_k}{A_k}Z{\Delta ^{\frac{1}{2}}} + \lambda _k^2R_k^2 - 2{A_k}Z\Delta  - {\lambda _k}{R_k}{\Delta ^{\frac{1}{2}}}\Bigg) \nonumber\\
&>\frac{L}{{{{R_k\left( {1 - {\rho _k}} \right)}^2}A_k^4{Z^2}\Delta }} \cdot \Bigg[4\gamma {\lambda _u}L\lambda _k^2R_k^2{Z^2}A_k^3 \nonumber \\
&\;\;\;\;\;\;\;\;\;\;\;\;\;\;\;\;  + {\lambda _k}{R_k}\left( {2{A_k}Z{\Delta ^{\frac{1}{2}}} + {\lambda _k}{R_k} - {\Delta ^{\frac{1}{2}}}} \right)\Bigg],
\end{align}
where $\Delta  = \lambda _k^2R_k^2 + 4\gamma {\lambda _\textrm{u}}{\lambda _k}{R_k} A_k^2 LZ$.
Since $\Delta^{\frac{1}{2}}>{\lambda _k}{R_k} $,
we further have
\begin{align}
&\frac{{{d^2}{{\bar D}_k}}}{{d{A_k}^2}}>\frac{L}{{{{R_k\left( {1 - {\rho _k}} \right)}^2}A_k^4{Z^2}\Delta }} \cdot \Bigg[4\gamma {\lambda _u}L\lambda _k^2R_k^2{Z^2}A_k^3 \nonumber \\
&\;\;\;\;\;\;\;\;\;\;\;\;\;\;\;\; + {\lambda _k}{R_k}\left( {2{A_k}Z{\Delta ^{\frac{1}{2}}} + {\lambda _k}{R_k} - {\Delta ^{\frac{1}{2}}}} \right)\Bigg] \nonumber
\end{align}
\begin{align}\label{convex3}
&> \frac{L}{{{{R_k\left( {1 - {\rho _k}} \right)}^2}A_k^4{Z^2}\Delta }} \cdot \Bigg[4\gamma {\lambda _u}L\lambda _k^2R_k^2{Z^2}A_k^3 \nonumber \\
&\;\;\;\;\;\;\;\;\;\;\;\;\;\;\;\; + {\lambda _k}{R_k}\left( {2{\lambda _k}{R_k}{A_k}Z + {\lambda _k}{R_k} - {\Delta ^{\frac{1}{2}}}} \right)\Bigg] \nonumber \\
&\mathop  > \limits^{\left( a \right)} \frac{{4\gamma {\lambda _u}L^2\lambda _k^2R_k}}{{{{\left( {1 - {\rho _k}} \right)}^2}{A_k}\Delta }}>0,
\end{align}
where (a) follows from the fact that $\rho_k<1$,
As the constraints \eqref{Constraints1p} and \eqref{Constraints2p} are linear, it can be concluded from \eqref{convex3} that the optimization problem is convex with respect to $A_k$.
\end{proof}

\section{Proof of Lemma 4}
\begin{proof}
We divide the proof into two parts.
\vspace{6pt}

\noindent 1) If $\gamma >\mathop {\max }\limits_{\forall k} \left\{ {\frac{{\left( {Z {+} 1} \right){\lambda _k}{R_k}}}{{{\lambda _\textrm{u}}L}}} \right\}$, then the mean queuing delay $\bar D_{k}$ of all tiers go to infinity as $A_k$ approaches to 1. Therefore, according to \eqref{DelayTotal}, the lower bound of the network mean queuing delay, $\bar D$, goes to infinity at the boundary of $\textbf{A}$. As $\bar D$ is convex within the region $\textbf{A}$, \eqref{Patial} always has a unique solution of the optimal association probabilities $\{A^*_k\} _{\forall k}$.

\noindent 2) If $\gamma <\mathop {\max }\limits_{\forall k} \left\{ {\frac{{\left( {Z {+} 1} \right){\lambda _k}{R_k}}}{{{\lambda _\textrm{u}}L}}} \right\}$, then there exists at least one tier such that the lower bound of its mean queuing delay is always bounded.
Without loss of generality, denote this tier as Tier $K$. For Tier $K$, we have $\frac{{{\lambda _K}{R_K}}}{{\gamma {\lambda _\textrm{u}}L {-} {\lambda _K}{R_K}Z}}>1$, and the feasible region $\textbf{A}$ is then written as
\begin{align}
\textbf{A}{=}\Bigg\{\left( {{A_1},...,{A_{K{-}1}}} \right),\left| 0<{A_k} < \min\Big\{1,\frac{{{\lambda _k}{R_k}}}{{\gamma {\lambda _\textrm{u}}L {-} {\lambda _k}{R_k}Z}}\Big\}, \right.\nonumber
\end{align}
\begin{align}\label{region2}
k \in {\{1...,K {-} 1} \};\; 0<\sum\limits_{k = 1}^{K {-} 1} {{A_k}}<1 \Bigg\}.
\end{align}
For each $k\in\{1,\ldots,K{-}1\}$, we have
\begin{align}
&\mathop {\lim }\limits_{{A_k} \to 0} \frac{{\partial \bar D}}{{\partial {A_k}}}= 2{\lambda _K}Z\nonumber \cdot
\end{align}
\begin{align}\label{Patial2}
\frac{{R_K}A_K^{ - 2}\left[1-{{\left( {1 {+} 4\gamma {\lambda _\textrm{u}}\lambda _K^{ - 1}{A^2_K}{R^{-1}_K}LZ} \right)}^{ - \frac{1}{2}}}\right]}{{{{\left[ {2Z{R_K} {+} {R_K}A_K^{ - 1} {-} {{\left( {R_K^2A_K^{ - 2} {+} 4\gamma {\lambda _\textrm{u}}\lambda _k^{ - 1}{R_k}LZ} \right)}^{\frac{1}{2}}}} \right]}^2}}}<0
\end{align}
according to \eqref{Patial}.

Following a similar approach, if $\frac{{{\lambda _k}{R_k}}}{{\gamma {\lambda _\textrm{u}}L {-} {\lambda _k}{R_k}Z}}>1$, we have
\begin{equation}\label{right1}
\mathop {\lim }\limits_{{A_k} \to 1} \frac{{\partial \bar D}}{{\partial {A_k}}} > 0.
\end{equation}
Otherwise, if $\frac{{{\lambda _k}{R_k}}}{{\gamma {\lambda _\textrm{u}}L {-} {\lambda _k}{R_k}Z}}<1$, the lower bound $\bar D_{k}$ goes to infinity as $A_k$ approaches $\frac{{{\lambda _k}{R_k}}}{{\gamma {\lambda _\textrm{u}}L {-} {\lambda _k}{R_k}Z}}$, and thus we have
\begin{equation}\label{right2}
\mathop {\lim }\limits_{{A_k} \to \frac{{{\lambda _k}{R_k}}}{{\gamma {\lambda _u}L {-} {\lambda _k}{R_k}Z}}} \frac{{\partial \bar D}}{{\partial {A_k}}} > 0.
\end{equation}
By combining \eqref{Patial2}-\eqref{right2}, it can be concluded that \eqref{Patial} always has only one solution within the region $0<{A_k} < \min\{1,\frac{{{\lambda _k}{R_k}}}{{\gamma {\lambda _\textrm{u}}L {-} {\lambda _k}{R_k}Z}}\}$, $k\in\{1,\ldots,K{-}1\}$.

Furthermore, if $\sum\limits_{k = 1}^{K {-} 1} {{A_k}}>1$, i.e., $A_K<0$, we always have $\frac{{\partial \bar D}}{{\partial {A_k}}} > 0$, $k\in\{1,\ldots,K{-}1\}$ by substituting $A_K<0$ into \eqref{Patial}. This indicates that the solution is not in the region where $\sum\limits_{k = 1}^{K {-} 1} {{A_k}}>1$. Therefore, \eqref{Patial} has a unique solution in region $\textbf{A}$ when $\gamma <\mathop {\max }\limits_{\forall k} \left\{ {\frac{{\left( {Z {+} 1} \right){\lambda _k}{R_k}}}{{{\lambda _\textrm{u}}L}}} \right\}$.
\end{proof}

\section{Derivation of \eqref{Optimal}}
By combining \eqref{rho_app}, \eqref{DelayTotal}, and \eqref{Constraints1p}, when the mean packet arrival rate of each user satisfies $\gamma <\mathop {\min }\limits_{\forall k} \left\{ {\frac{{\left( {Z {+} 1} \right){\lambda _k}{R_k}}}{{{\lambda _\textrm{u}}L}}} \right\}$, the lower bound of the network mean queuing delay can be written as
\begin{align}\label{DelayT2app}
&\bar D=\frac{1}{\sum\limits_{j = 1}^K {{\lambda _j}} }\sum\limits_{k = 1}^K {\frac{{\lambda _k^2L}}{{{\lambda _k}R_k{-} \gamma {\lambda _\textrm{u}}L{A_k}}}}  =\frac{1}{\sum\limits_{j = 1}^K {{\lambda _j}} }\Bigg[ \nonumber \\ &\sum\limits_{k = 1}^{K{-}1} \frac{{\lambda _k^2L}}{{{\lambda _k}R_k {-} \gamma {\lambda _\textrm{u}}L{A_k}}} + \frac{{\lambda _K^2L}}{{{\lambda _K}R_K {-} \gamma {\lambda _\textrm{u}}L({1 {-} \sum\limits_{j = 1}^{K {-} 1} {{A_j}} })}}\Bigg]
\end{align}
where $R_k$ is given by \eqref{Rate}.
By setting the partial derivative of $\bar D$ with respect to $A_k$ to zero, we have
\begin{align}
\frac{{\partial \bar D}}{{\partial {A_k}}} = \frac{{\lambda _k^2}}{{\sum\limits_{j = 1}^K {{\lambda _j}} }} \cdot \frac{{{\lambda _\textrm{u}}\gamma }}{{{{\left( {{\lambda _k}\frac{{{R_k}}}{L} - {\lambda _\textrm{u}}\gamma A_k} \right)}^2}}}- \frac{{\lambda _K^2}}{{\sum\limits_{j = 1}^K {{\lambda _j}} }}\cdot \nonumber
\end{align}
\begin{align}\label{Derivative}
 \frac{{{\lambda _\textrm{u}}\gamma }}{{{{\left[ {{\lambda _k}\frac{{{R_k}}}{L} - {\lambda _\textrm{u}}\gamma \left( {1 - \sum\limits_{j = 1}^{K - 1} {{A_j}} } \right)} \right]}^2}}}= 0,\;\forall k{\in}\{1,\ldots,K{-}1\}.
\end{align}
By combining \eqref{Constraints1p} and \eqref{Derivative}, \eqref{Optimal} can be obtained.
\end{appendices}

\end{document}